\documentclass[aps,epsfig,preprint]{revtex4-1}
\usepackage[section]{placeins}        
\usepackage{float}
 \usepackage{graphics}
 \usepackage{multirow}
 \usepackage{soul}
\usepackage{caption}
\usepackage{epsfig}
\usepackage{pdfpages}
\usepackage{subcaption}\usepackage{ragged2e}
\usepackage{amsmath,amsthm, amssymb, latexsym}

\newcommand{\be}{\begin{equation}}
\newcommand{\ee}{\end{equation}}

\newcommand{\bea}{\begin{eqnarray}}
\newcommand{\eea}{\end{eqnarray}}

\begin{document}
\title{2-D studies of Relativistic electron beam plasma instabilities  in an inhomogeneous plasma}
 \author{Chandrashekhar Shukla, Amita Das}
 \email{amita@ipr.res.in}
 \affiliation{Institute for Plasma Research, Bhat , Gandhinagar - 382428, India }
  \author{Kartik Patel} 
 \affiliation{Bhabha Atomic Research Centre, Trombay, Mumbai - 400 085, India }
\date{\today}
\begin{abstract} 
Relativistic electron beam propagation in plasma is fraught with several micro instabilities like two stream, filamentation etc., in plasma.  
This results in severe limitation of the electron transport through a 
plasma medium. Recently, however, there has been an experimental demonstration of  improved  transport of Mega Ampere of electron currents (generated by the interaction of intense laser with solid target) in a carbon nanotube structured solid target [Phys. Rev Letts. {\bf{108}}, 235005 (2012)]. This then  suggests that the 
inhomogeneous plasma (created by the ionization of carbon nano tube structured target) helps in containing the growth of
 the beam plasma instabilities. This manuscript addresses this issue with the help of a detailed analytical study and simulations 
with the help of 2-D Particle - In - Cell code. The study conclusively 
 demonstrates that the growth rate of the dominant instability in the 2-D geometry   
 decreases when the plasma density is chosen to be inhomogeneous,  
 provided the scale length $1/k_{s}$ of the inhomogeneous plasma  is less than the typical plasma skin depth ($c/\omega_{p0}$) scale.
At such small scale lengths channelization of currents are also observed in simulation. 
\end{abstract}
 \maketitle 
\section{Introduction}
The relativistic beam plasma system occurs in the context of many frontier research areas in both laboratory and space plasmas, 
such as the fast ignition scheme of inertial 
 confinement fusion \cite{tabak_05}, compact particle accelerators \cite{malka}, solar flares physics \cite{solarflare}, cosmic magnetic field generation \cite{cosmaggen} and magnetic field reconnection \cite{s_bulnov}. Therefore, the 
 study of instabilities associated with the relativistic beam-plasma systems is  of central importance in understanding issues pertaining to 
 a better transport of   
 beam electrons in the plasma and provide clues  to  improve upon it.  The manuscript explores this in detail.  
  
 When a short, intense laser pulse (I$\geq 10^{19}$W/$cm^{2}$) irradiates a solid target, a relativistic electron beam is generated through the process of wave breaking \cite{Modena,malka,joshi}. 
 This relativistic electron beam can propagate inside the plasma target even when the  current associated with it 
 exceeds the Alfven current limit $I= (mc^{3}/e)\gamma = 17{\gamma_{b}}$ kA,  here $m$ is the electron mass, $e$ is the electronic charge, $c$ is speed of light and $\gamma_{b}$ is the Lorentz factor of the beam. This is because simultaneously 
a return current is induced by the electrons of the background plasma which  compensates the forward 
beam electron current. This combination of forward  and return current is, however,  highly unstable to two stream and 
filamentation instabilities. For the two stream  (longitudinal mode) \cite{bhom} instability, the 
perturbations  propagate parallel to the   direction of the beam flow. 
When the perturbations propagate normal to the beam propagation direction they are  termed 
as the filamentation instability (transverse mode) \cite{fried,lee}.  
This is also often known as the Weibel instability \cite{weibel}.

In the general case  the perturbation  would  propagate in a direction oblique to the beam and would 
have both  filamentation and two stream characteristics. 
 The instabilities associated with the  obliquely propagating  perturbations 
  have been  investigated  in the cold fluid limit \cite{Taguchi} as well as by  kinetic approaches \cite{kinetic}. 
The  particle-in-cell (PIC) simulations \cite{Dieckmann,shukla} have been carried out to validate the results. It has been shown  that when  $\gamma_{b}>1$, relativistic effects come in to play and the filamentation instability dominates over the two-stream instability \cite{bret}. 
A full hierarchical map showing the parameter regions (ratio of the beam density to the plasma density and $\gamma_{b}$) 
where these modes dominate has been provided in  \cite{review}. The maximally growing mode amongst the 
 spectrum of  unstable modes  determines the beam evolution and sets the stage for the 
 subsequent non-linear evolution. 
 In fact the magnetic field generation is also closely linked  with these modes and occurs 
 due to the current imbalance created by the most dominant  modes of the instability. 
 As the amplitude of the perturbations increase the nonlinear coupling defines the spectral cascade to other 
 scales making the magnetic power spectrum turbulent in many scenarios.  This in turn is detrimental to beam transport.

There are many attempts at avoiding and/or suppressing such instabilities with are responsible for current separation.  For instance, 
 theoretical and experimental work on collimated propagation of electron beam assisted by resistivity gradient have been reported in Ref. \cite{A.R.BELL,B.RAMA}. 
 A recent experiment \cite{g.r.k} has demonstrated  transport of Mega Amperes of electron currents through  nano structured target  over distances of the order of  millimeter. The transportation of electron
 beam  over such  distances is  possible only if the beam plasma associated instabilities are somehow suppressed 
 and the separation between forward and return currents does not take place as rapidly as predicted by the growth rate of 
 these instabilities.
 In a simplified 1-D treatment, the suppression of Weibel/filamentation instability by the 
 equilibrium density inhomogeneity (due to the nanostructures) has been shown in some recent studies \cite{mishra,chandra}. 
 However, for the  parameters of experimental system \cite{g.r.k}, typically the oblique mode would be the fastest growing mode \cite{review}. 
This necessitates a  2-D treatment wherein the role of plasma inhomogeneity on the growth of oblique mode should be investigated. 
Such a study is the main objective of the work presented in this manuscript. 
 
We have carried out the linear analysis of the beam plasma instability 
in an inhomogeneous plasma medium by using the coupled  Maxwell and two-fluid electron equations in the  relativistic regime. The inhomogeneity in the beam-plasma density has been chosen to have  a sinusoidal profile orthogonal to the direction of the beam propagation.  
 This inhomogeneous density mimics the scenario which would arise when a high power laser of intensity (I$\geq 10^{19}$ W/$cm^{2}$)  incident on a structured  target ionizes it.  The propagation of the accelerated relativistic electron beam (REB) 
occurs at a very fast time scale at which the heavier ions are unable to respond and continue to 
 maintain the inhomogeneous profile of their density. Such a inhomogeneous background ion density constrains the 
flow of the beam and background plasma electrons. 
The linear analysis of the coupled fluid Maxwell set of equation shows  that when the plasma  density ripples 
have a scale length less than the skin depth,  the dominant oblique mode gets suppressed. We have also 
carried out a 2D PIC simulations which corroborates with the linear results. With the help of PIC 
simulations one has also been able to explore the nonlinear regime of the system. It is observed that  
 magnetic field energy is comparatively lower  for the inhomogeneous cases in the nonlinear regime. This 
 further  suggests that the current separation is weak when the plasma is inhomogeneous, conducive to  the 
 uninhibited propagation of the REBs.
 The manuscript has been organized as follows. The theoretical model (coupled Maxwell and two electron fluid) and 
 governing equation for linear analysis  for the beam -plasma system has been presented in section II. The 
 growth rate of the unstable mode has been analyzed   in section III.
 In section IV, the observations from  PIC simulation have been presented. A comparison between  
  homogeneous and inhomogeneous density plasma is also provided which shows that 
  if the scale length of the inhomogeneity is sharper than the electron skin depth the growth of beam plasma instability  
  gets suppressed. In section V, we present our conclusions.

\section{Theoretical Model And Governing Equations }
\begin{figure}[1]
                \includegraphics[width=\textwidth]{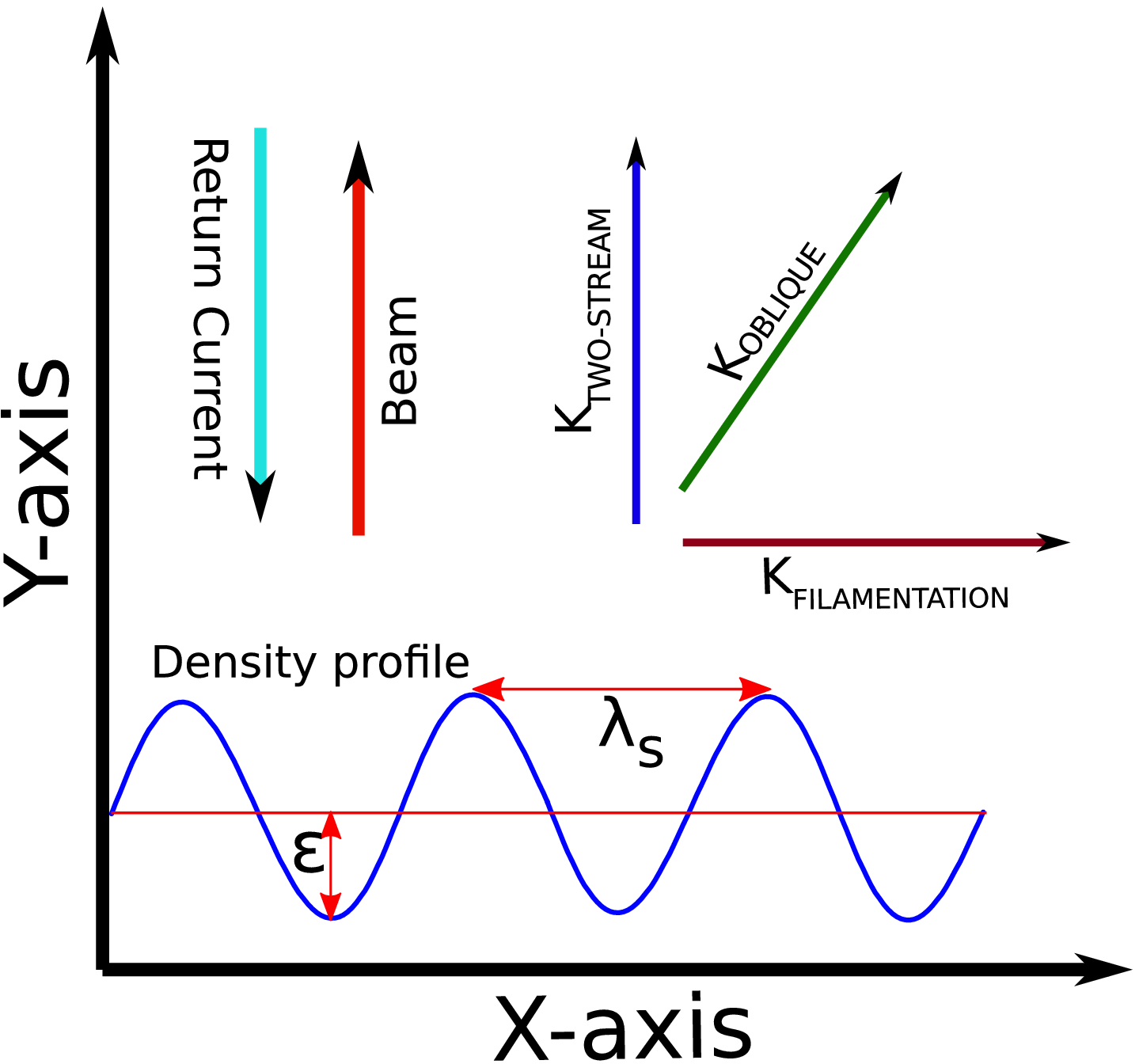} 
              \justifying   \caption{ Sketch of the system considered in the present article. }  
                 \label{fig:2ddensity_profile}
         \end{figure}

We employ a two fluid description wherein the relativistic electron beam (REB) and the background electrons are 
treated as separate fluids. The evolution equations of these two electron species are coupled with the Maxwell's 
equations which governs the field evolution.  
The ions are assumed to be static and merely provide a neutralizing background. 
  A collision-less, quasi-neutral and  un-magnetized electron beam-plasma system is considered. 
  The  electron beam drift in the \^{y} direction inducing a 
return current in cold background plasma electrons ($T_{0p}$). 
 The  initial configuration is chosen so as to have  charge and
current neutrality. The sketch of the configuration is shown in Fig.~\ref{fig:2ddensity_profile}. 
At time t=0, a space dependent (transverse to the beam flow direction) background ion density $n_{0i}$ is chosen. 
The charge neutrality condition is fulfilled by choosing the sum of the beam density $n_{0b}$ and the background 
electron density $n _{0p}$ to match with background density of ions.
The system can be described by following set of dimensionless governing equations:
 \begin{eqnarray}
&&\frac{\partial n_{\alpha}}{\partial t}  +  \nabla \cdot\left(n_{\alpha}\vec{v_{\alpha}}\right) 
  = 0 \\
\label{con1}
&&\frac{\partial \vec{p_{\alpha}}}{\partial t} + \vec{v_{\alpha}} \cdot \nabla \vec{p_{\alpha}}
  =-\left(\vec{E}+\vec{v_{\alpha}}\times \vec{B}\right) -\frac{\nabla P_{\alpha}}{n_{\alpha}} \\
 \label{mom}
&&\frac{\partial \vec{B}}{\partial t}  
  = -\nabla \times \vec{E}\\
\label{Max2}
&&\frac{\partial \vec{E}}{\partial t}  
  = \nabla \times \vec{B}- \Sigma_{\alpha}{\vec{J_{\alpha}}}
\label{Max3}
\end{eqnarray}
with
$\vec v_{\alpha}={\vec{p_{\alpha}}}/{(1+{p_{\alpha}}^{2})^{\frac{1}{2}}}$ where $\vec{p_{\alpha}}$ is momentum vector,  $\vec{J_{\alpha}}=-n_{\alpha}\vec{v_{\alpha}}$. 
The pressure $P_{\alpha}$ is provided by the equation of state. In the above equations, velocity is normalized by 
the speed of light $c$, density by $n_{0}$ (the spatially average ion density), frequency by  $\omega_{0}=\sqrt{4\pi n_{0}^2e/m_{e}}$, length by electron skin depth $d_e=c/\omega_0e$ and electric and
magnetic field by $E_{0}=B_{0}=m_{e}c\omega_{0}/e$ where $m_{e}$ is electron rest mass and $e$ is electron charge. The subscript $\alpha$ is $'b'$ for beam and $'p'$ for plasma.
 
In the equilibrium there is no electric and magnetic field, so there is complete charge as well as current neutralization. 
This is achieved by balancing the forward and return electron currents at each spatial location. 
  For simplicity of treatment, the profile of transverse temperature $T_{\alpha}$ is chosen in 
such a way that gradient of pressure is zero in equilibrium. 
The temperature parallel to beam propagation direction has been  chosen to be negligible. The background plasma is chosen to be  
cold ($T_{0p}=0$) in all our analytical as well  simulation studies. Thus, for an inhomogeneous case considered by 
us in this work  we have the following conditions for 
equilibrium: 
 \begin{equation} 
n_{0i}(x) 
  =n_{0b}(x)+n_{0p}(x)
\label{quasi}
\end{equation}
\begin{equation} 
 \Sigma_{\alpha} n_{0 \alpha}(x)\vec{v_{0 \alpha}}=0
\label{null-current}
\end{equation}
 In equilibrium beam pressure $P_{0b}$ is chosen to be independent of x. This is achieved by choosing the beam temperature $T_{0b}(x)$ to be satisfying the following   condition
\begin{equation} 
P_{0b}=T_{0b}(x) n_{0b}(x)=constant=C.
\end{equation}
\begin{equation} 
 T_{0b}(x)=C/n_{0b}(x)
\label{pressure_balance}
\end{equation}
 The suffix $0$ indicates the equilibrium fields. We linearize equations (1)-(4) to obtain linear growth rate 
 of the instability. The inhomogeneity being periodic along $'x'$ we choose 
 all the perturbed quantities to have the  form of 
 $f_{\alpha}=\Sigma_{j}^{....,\pm2,\pm1,0}f_{\alpha j}e^{i((k_{x}+jk_{s})x+k_{y}y-\omega t)}$. Here $k_s$ is the wavenumber associated with the periodic 
 profile of inhomogeneity.
 For small amplitude of inhomogeneity$(\varepsilon<1)$,  the terms corresponding to $j=0,\pm 1$ are the ones which are only retained.  
 All higher order terms are neglected.
 The coupled set of differential equations obtained after  linearizing Eqs.(1)-(4)
are following : 
   \begin{eqnarray}
&&\Omega_{\alpha}\gamma_{0\alpha}^3 v_{l\alpha y}'' + \left(\omega^2-k_{y}^2\right)\Omega_{\alpha}\gamma_{0\alpha}^3 v_{l\alpha y}+
\left(\omega^2-k_{y}^2\right)\sum_{\alpha}\frac{n_{0\alpha} }{\Omega_{\alpha}}v_{l\alpha y}+\sum_{\alpha}\frac{k_{y}-\omega v_{0\alpha}}{\Omega_{\alpha}}\left(n_{0\alpha} v_{l\alpha y}\right)' =0\\ 
 \label{vy_in} 
&&\hspace{0.8cm}\frac{(\omega^2-k_{y}^2)}{\Omega_{\alpha}}\eta T_{0\alpha}v_{l\alpha x}''-\frac{(\omega^2-k_{y}^2)}{\Omega_{\alpha}}\eta T_{0\alpha}'v_{l\alpha x}' \nonumber
+\left(\frac{(\omega^2-k_{y}^2)}{n_{0\alpha}\Omega_{\alpha}}\eta T_{0\alpha}'n_{0\alpha}''+(\omega^2-k_{y}^2)\Omega_{\alpha}\gamma_{0\alpha}\right)v_{l\alpha x}\\
 &&\hspace{0.8cm}-\Omega_{\alpha}\sum_{\alpha}n_{0\alpha}v_{l\alpha x} +\left(i\frac{k_{y}}{\Omega_{\alpha}}(\omega^2-k_{y}^2)\eta T_{0\alpha} - i\left(k_{y}-v_{0\alpha}\omega\right)\Omega_{\alpha}\gamma^3_{0\alpha}\right)v_{l\alpha y}' =0 \\ \nonumber
\label{vx_in}
   \end{eqnarray}
   Where $\Omega_{\alpha}$=$(\omega-k_y v_0\alpha)$ and $\eta$ is ratio of specific heat.
 When the plasma density is homogeneous i.e. $\varepsilon = 0$,  these coupled equations reduce to the  standard 
 linear  equations for  the beam plasma system: 
  \begin{eqnarray}
 &&\Omega_{\alpha}\gamma_{0\alpha}^3 v_{l\alpha y}'' + \left(\omega^2-k_{y}^2\right)\Omega_{\alpha}\gamma_{0\alpha}^3 v_{l\alpha y}+
\left(\omega^2-k_{y}^2\right)\sum_{\alpha}\frac{n_{0\alpha} }{\Omega_{\alpha}}v_{l\alpha y}+\sum_{\alpha}\frac{k_{y}-\omega v_{0\alpha}}{\Omega_{\alpha}}n_{0\alpha} v_{l\alpha y}' =0  \\ 
 \label{hvy_in}
&&\frac{(\omega^2-k_{y}^2)}{\Omega_{\alpha}}\eta T_{0\alpha}v_{l\alpha x}'' \nonumber
+\left(\omega^2-k_{y}^2\right)\Omega_{\alpha}\gamma_{0\alpha}v_{l\alpha x} -\Omega_{\alpha}\sum_{\alpha}n_{0\alpha}v_{l\alpha x}\\ 
&&\hspace{05.4cm}+\left(i\frac{k_{y}}{\Omega_{\alpha}}(\omega^2-k_{y}^2)\eta T_{0\alpha} - i\left(k_{y}-v_{0\alpha}\omega\right)\Omega_{\alpha}\gamma^3_{0\alpha}\right)v_{l\alpha y}' =0 \\ \nonumber
\label{hvx_in}
   \end{eqnarray}
   In the next section we study in detail the growth rate of the most unstable mode for the general case by using eq.~(\ref{vy_in}).
   \section{Analytical Study }
 The linear evolution of beam-plasma system is studied by choosing sinusoidal form of ion background density 
of the  form 
 $n_{0i}=n_{0}(1+\varepsilon cos(k_{s}x))$ where $n_{0}$ is constant normalized density, $\varepsilon$ is 
 inhomogeneity amplitude and $k_{s}= 2\pi m/L_x$ ($m$ is an integer and $L_x$ is the system dimension along $x$) is inhomogeneity wave vector.
 The equilibrium beam and plasma density profile have been taken as 
 $n_{0b}=\beta_{b} n_{0}(1+\varepsilon_{b} cos(k_{s}x))$, $n_{0p}=\beta_{p} n_{0}(1+\varepsilon_{p} cos(k_{s}x))$ to maintain quasi neutrality where $\sum_{\alpha}\beta_{\alpha}=1$ is a fraction. For simplicity, 
 we have chosen $\varepsilon_{b}=\varepsilon_{p}=\varepsilon$ here.

  To evaluate the  linear growth rate of oblique mode driven instability, we expand eq.~(\ref{vy_in}) and eq.~(\ref{vx_in}) and solve it by substituting for the Bloch wave function form of the perturbed fields in the periodic system. Retaining only the first order terms, as mentioned earlier 
  we obtain: 
    \begin{eqnarray}
     &&\Omega_{\alpha}\gamma_{0\alpha}^3 \left(\lambda^2-k_{d}^2\right)v_{d\alpha y}+ \lambda^2\sum_{\alpha}\frac{\beta_{\alpha}}{\Omega_{\alpha}}v_{d\alpha y}+\nonumber
\lambda^2\sum_{\alpha}\frac{0.5\varepsilon\beta_{\alpha}}{\Omega_{\alpha}}v_{\alpha y} +i\sum_{\alpha}\frac{\zeta_{\alpha}}{\Omega_{\alpha}}\beta_{\alpha}k_{d}v_{d\alpha x}\\
&&\hspace{10.2cm}+i\sum_{\alpha}\frac{\zeta_{\alpha}}{\Omega_{\alpha}}0.5\varepsilon\beta_{\alpha}k_{d}v_{\alpha x} =0 \\\nonumber
 &&\Omega_{\alpha}\gamma_{0\alpha}^3 \left(\lambda^2-k_{o}^2\right)v_{\alpha y}+ \lambda^2\sum_{\alpha}\frac{\beta_{\alpha}}{\Omega_{\alpha}}v_{\alpha y}+ \nonumber
\lambda^2\sum_{\alpha}\frac{0.5\varepsilon\beta_{\alpha}}{\Omega_{\alpha}}\left(v_{d\alpha y}+v_{a\alpha y} \right) 
+i\sum_{\alpha}\frac{\zeta_{\alpha}}{\Omega_{\alpha}}\beta_{\alpha}k_{o}v_{\alpha x}\\&&\hspace{8.0cm}+i\sum_{\alpha}\frac{\zeta_{\alpha}}{\Omega_{\alpha}}0.5\varepsilon\beta_{\alpha}k_{o}\left(v_{d\alpha x}+v_{a\alpha x}\right) =0 \\\nonumber
 &&\Omega_{\alpha}\gamma_{0\alpha}^3 \left(\lambda^2-k_{a}^2\right)v_{a\alpha y}+ \lambda^2\sum_{\alpha}\frac{\beta_{\alpha}}{\Omega_{\alpha}}v_{a\alpha y}+\nonumber
\lambda^2\sum_{\alpha}\frac{0.5\varepsilon\beta_{\alpha}}{\Omega_{\alpha}}v_{\alpha y} 
+i\sum_{\alpha}\frac{\zeta_{\alpha}}{\Omega_{\alpha}}\beta_{\alpha}k_{a}v_{a\alpha x}\\&&\hspace{10.2cm}+i\sum_{\alpha}\frac{\zeta_{\alpha}}{\Omega_{\alpha}}0.5\varepsilon\beta_{\alpha}k_{a}v_{\alpha x} =0 \\ \nonumber
&&\left(\frac{\lambda^2}{\Omega_{\alpha}\beta_{\alpha}}\eta c_{\alpha}k_{d}^2-\lambda^2\Omega_{\alpha}\gamma_{0\alpha}\right)v_{d\alpha x}+
\Omega_{\alpha}\sum_{\alpha}\beta_{\alpha}v_{d\alpha x}+0.5\varepsilon\frac{\lambda^2}{\Omega_{\alpha}\beta_{\alpha}}\eta c_{\alpha}[k_{d}^2-k_{0}(2k_{0}-k_{s})]v_{\alpha x}\\ 
&&+0.5\varepsilon\Omega_{\alpha}\sum_{\alpha}\beta_{\alpha}v_{\alpha x}+\left(k_{y}k_{d}\frac{\lambda^2}{\Omega_{\alpha}\beta_{\alpha}}\eta c_{\alpha}-\zeta_{\alpha}\Omega_{\alpha}\gamma_{0\alpha}^3k_{d}\right)v_{d\alpha y}-
 0.5\varepsilon\frac{\lambda^2}{\Omega_{\alpha}\beta_{\alpha}}\eta c_{\alpha}k_{y}k_{0}v_{\alpha y} =0 \\\nonumber
&&\left(\frac{\lambda^2}{\Omega_{\alpha}\beta_{\alpha}}\eta c_{\alpha}k_{0}^2-\lambda^2\Omega_{\alpha}\gamma_{0\alpha}\right)v_{\alpha x}+
\Omega_{\alpha}\sum_{\alpha}\beta_{\alpha}v_{\alpha x}+0.5\varepsilon\frac{\lambda^2}{\Omega_{\alpha}\beta_{\alpha}}\eta c_{\alpha}[k_{0}^2-k_{d}(2k_{d}+k_{s})]v_{d\alpha x}+ \\ \nonumber
&& 0.5\varepsilon\frac{\lambda^2}{\Omega_{\alpha}\beta_{\alpha}}\eta c_{\alpha}[k_{0}^2-k_{a}(2k_{a}-k_{s})]v_{a\alpha x}+0.5\varepsilon\Omega_{\alpha}\sum_{\alpha}\beta_{\alpha}\left(v_{d\alpha x}+v_{a\alpha x}\right)\\ 
+&&\left(k_{y}k_{0}\frac{\lambda^2}{\Omega_{\alpha}\beta_{\alpha}}\eta c_{\alpha}-\zeta_{\alpha}\Omega_{\alpha}\gamma_{0\alpha}^3k_{0}\right)v_{\alpha y}-
 0.5\varepsilon\frac{\lambda^2}{\Omega_{\alpha}\beta_{\alpha}}\eta c_{\alpha}k_{y}\left(k_{d}v_{d\alpha y}+k_{a}v_{a\alpha y}\right) =0 \\ \nonumber
&&\left(\frac{\lambda^2}{\Omega_{\alpha}\beta_{\alpha}}\eta c_{\alpha}k_{a}^2-\lambda^2\Omega_{\alpha}\gamma_{0\alpha}\right)v_{a\alpha x}+
\Omega_{\alpha}\sum_{\alpha}\beta_{\alpha}v_{a\alpha x}+0.5\varepsilon\frac{\lambda^2}{\Omega_{\alpha}\beta_{\alpha}}\eta c_{\alpha}[k_{a}^2-k_{0}(2k_{0}+k_{s})]v_{\alpha x}\\ 
&&+0.5\varepsilon\Omega_{\alpha}\sum_{\alpha}\beta_{\alpha}v_{\alpha x}+\left(k_{y}k_{d}\frac{\lambda^2}{\Omega_{\alpha}\beta_{\alpha}}\eta c_{\alpha}-\zeta_{\alpha}\Omega_{\alpha}\gamma_{0\alpha}^3k_{a}\right)v_{a\alpha y}-
 0.5\varepsilon\frac{\lambda^2}{\Omega_{\alpha}\beta_{\alpha}}\eta c_{\alpha}k_{y}k_{0}v_{\alpha y} =0 \\ \nonumber
    \end{eqnarray}
where $\lambda^2=\left(\omega^2-k_{y}^2\right)$, $k_{d}=(k_x-k_s)$, $k_{0}=k_x$, $k_{a}=(k_x+k_s)$ and $\zeta_=(k_{y}-v_{0\alpha}\omega)$.
    
\begin{figure}[2]
                \includegraphics[width=\textwidth]{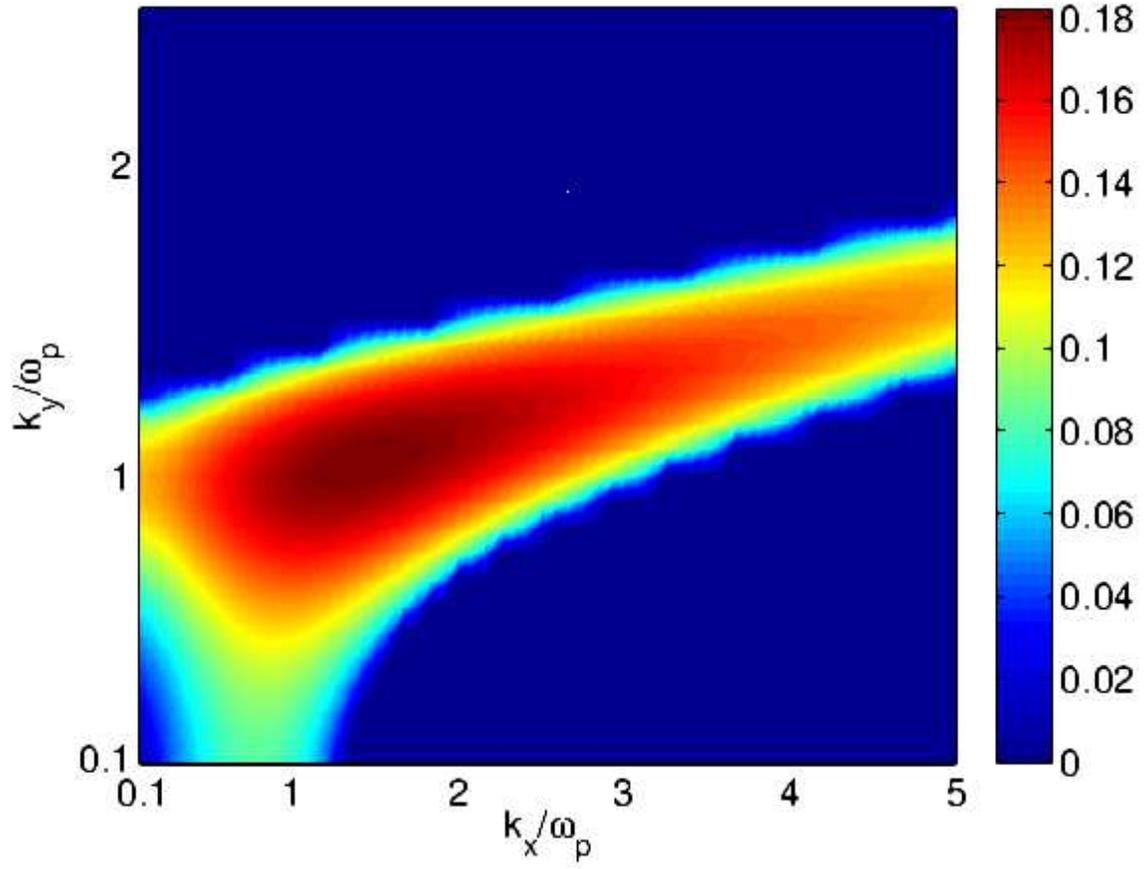} 
              \justifying   \caption{ Growth rate map of the 2D oblique instability for a homogeneous beam-plasma
                 at transverse beam temperature $T_{b0\perp}$=10 keV.  }  
                 \label{fig:hgrowth_rate}
         \end{figure}

 The map of the growth rate in the parameter space of $k_x$ and $k_y$ for the  homogeneous system has been shown in  Fig.~\ref{fig:hgrowth_rate} for   $n_{0b}/n_{0p}=1/9$ or $n_{0b}/n_{0e}=0.1$, $v_{0b}=0.9c$, $v_{0p}=-0.1c$ and $T_{0b\perp}=10 KeV$. 
 This  map shows that for this particular
set of parameters, oblique mode with($k_{x}$, $k_{y}$=(1.5,1)) has a maximum growth rate. Also a narrow oblique strip of 
 wavenumbers  extending up to $k_{x}$ =$\infty$ is found to be unstable.

\begin{figure}[3]
                \includegraphics[width=\textwidth]{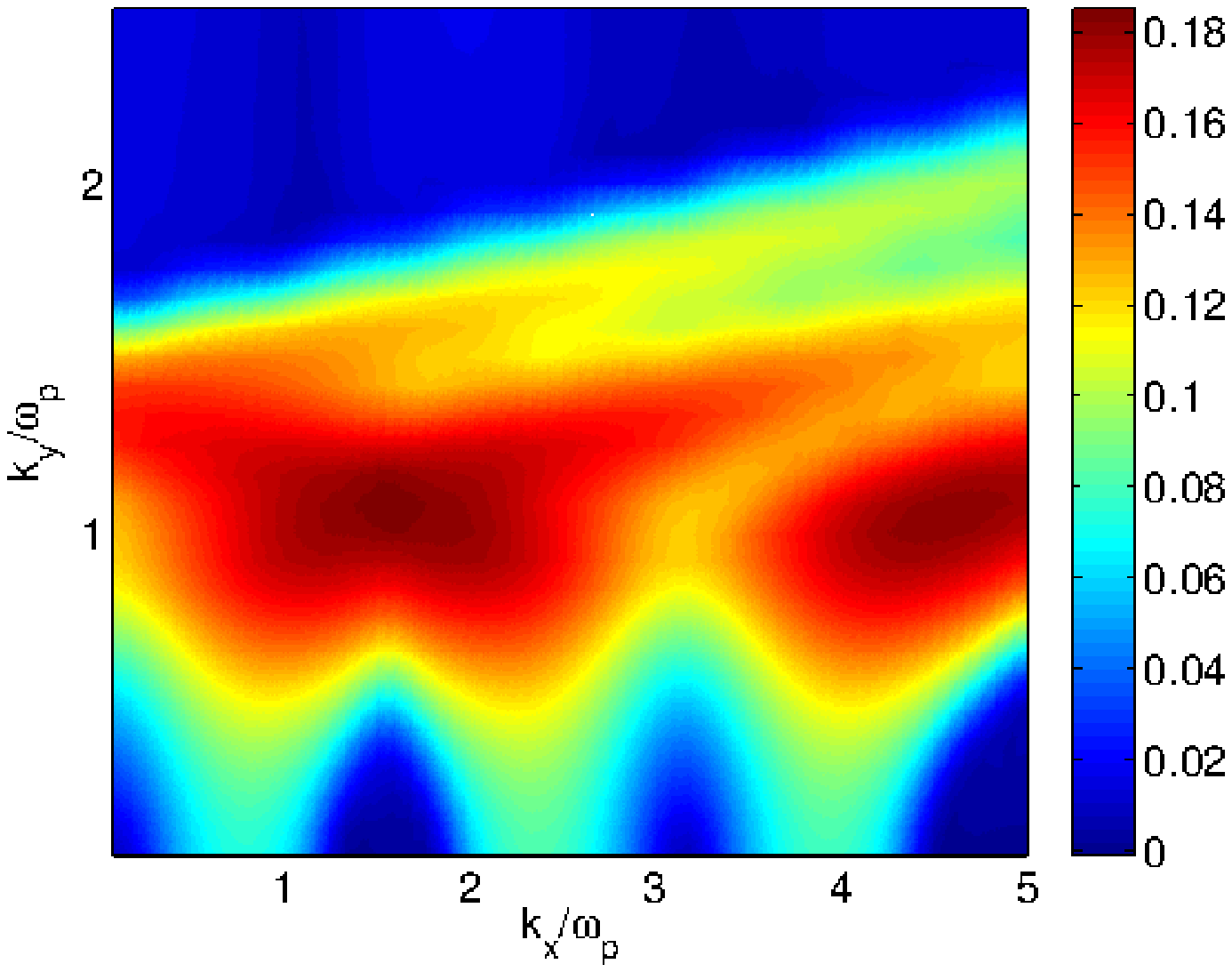}
              \justifying   \caption{ Growth rate map of the 2D oblique instability for a inhomogeneous beam-plasma
                 at transverse beam temperature $T_{b0\perp}$=10 keV at $k_{s}=\pi$
                 and $\varepsilon=0.1$.}  
                 \label{fig:in1}
         \end{figure}

\begin{figure}[4]
                \includegraphics[width=\textwidth]{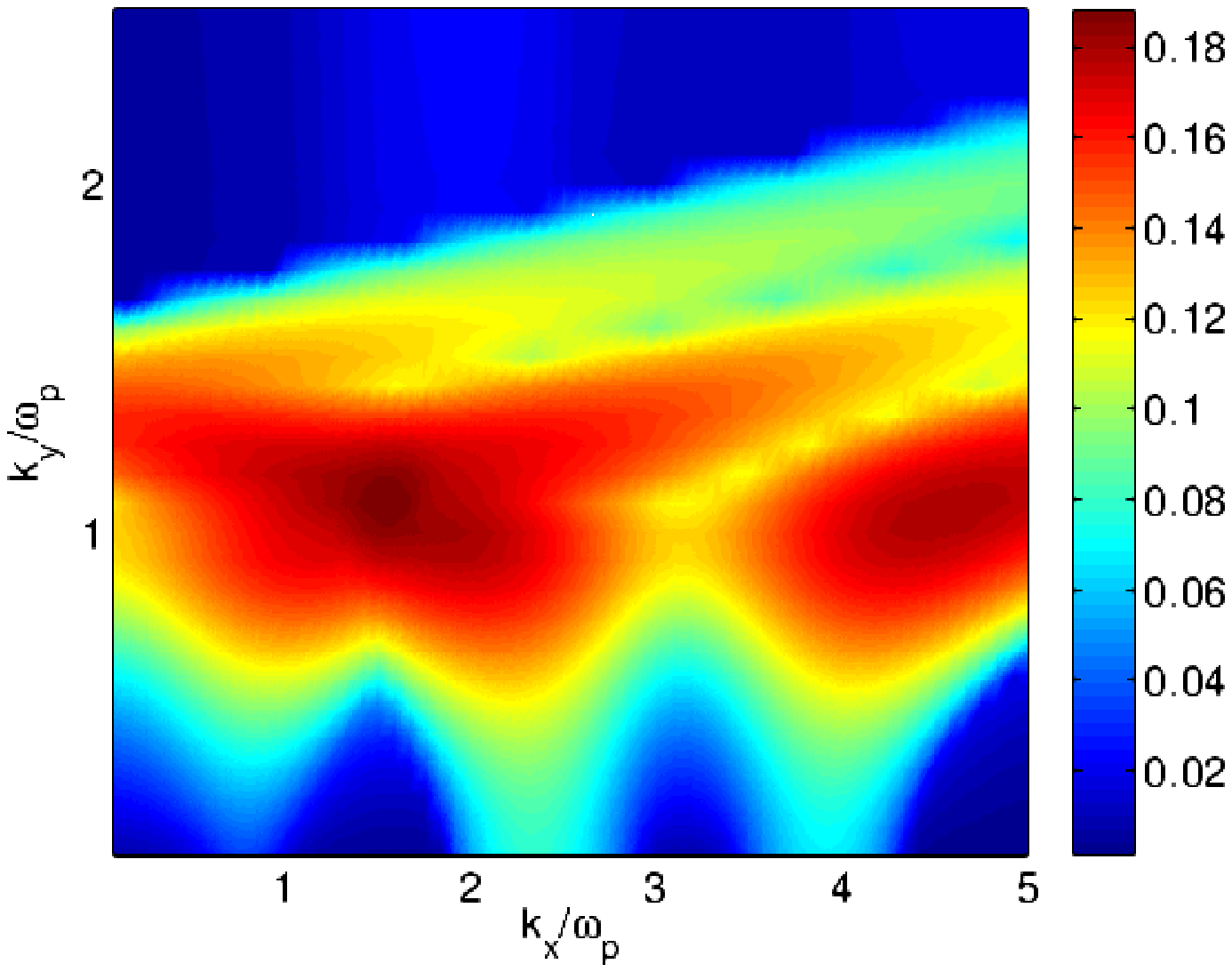} 
              \justifying   \caption{ Growth rate map of the 2D oblique mode driven instability for a inhomogeneous beam-plasma
                 at transverse beam temperature $T_{b0\perp}$=10 keV at $k_{s}=\pi$
                 and $\varepsilon=0.2$.}  
                 \label{fig:in2}
         \end{figure}

When the plasma density in chosen to be inhomogeneous and the scale length of the inhomogeneity is 
longer than the skin depth, i.e.    $ 2\pi/k_{s}>c/\omega_{0}$ the growth rate of oblique mode driven instability  is observed to increase with 
the inhomogeneity amplitude $\varepsilon$, as shown in Fig.~\ref{fig:in1} and ~\ref{fig:in2}. The unstable modes continues to stay 
around the  oblique patch in the  $k_x$ vs. $k_y$ plane. However, the growth rate is spread over a wider $k_y$ domain
compared to the homogeneous case. Another feature is the appearance of several maxima along $k_x$ in contrast to the single extrema for the homogeneous case.  

\begin{figure}[5]
                \includegraphics[width=\textwidth]{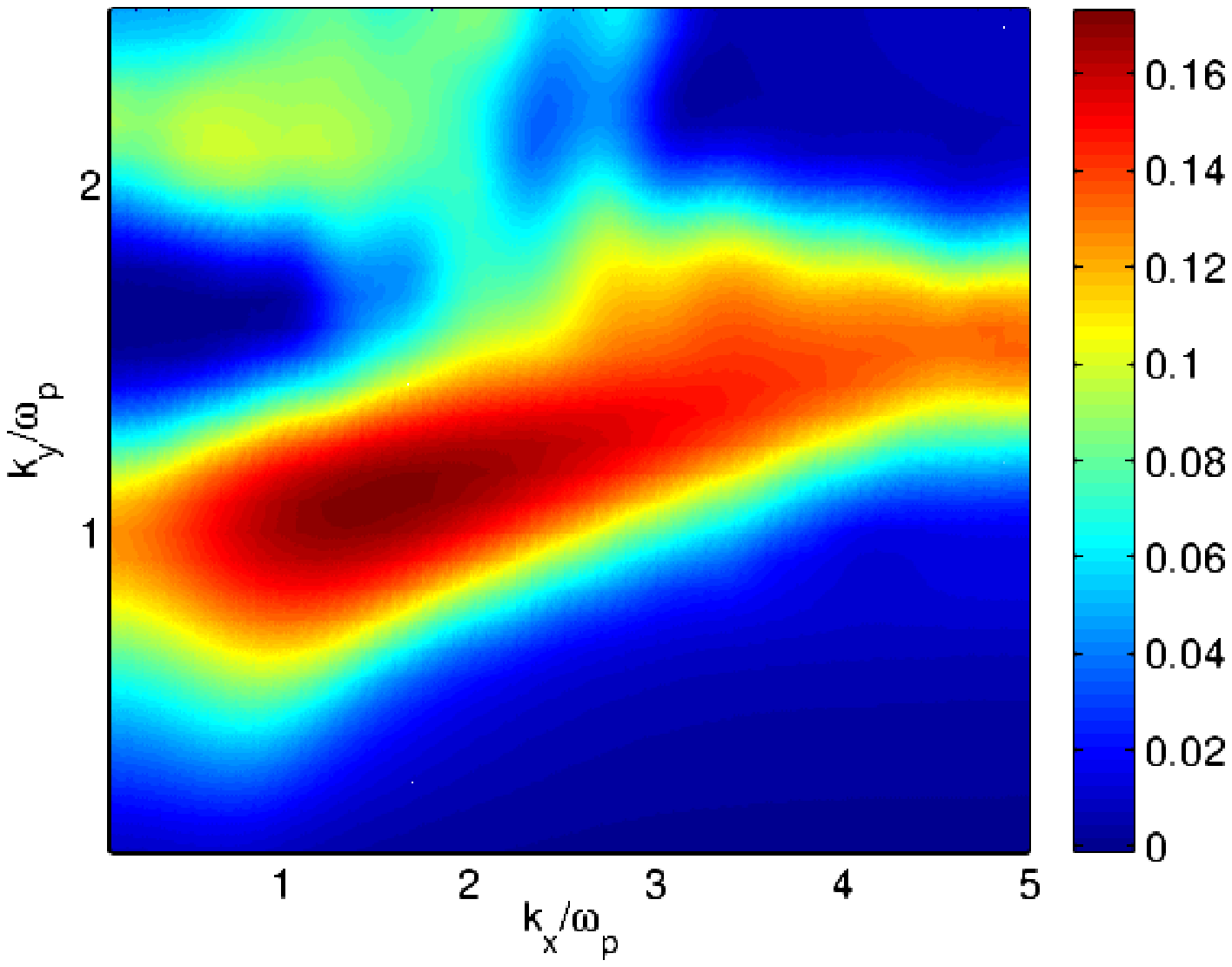}
              \justifying   \caption{ Growth rate map of the 2D oblique mode driven instability for a inhomogeneous beam-plasma
                 at transverse beam temperature $T_{b0\perp}$=10 keV at $k_{s}=3\pi$
                 and $\varepsilon=0.1$.}  
                 \label{fig:in3}
         \end{figure}

\begin{figure}[6]
                \includegraphics[width=\textwidth]{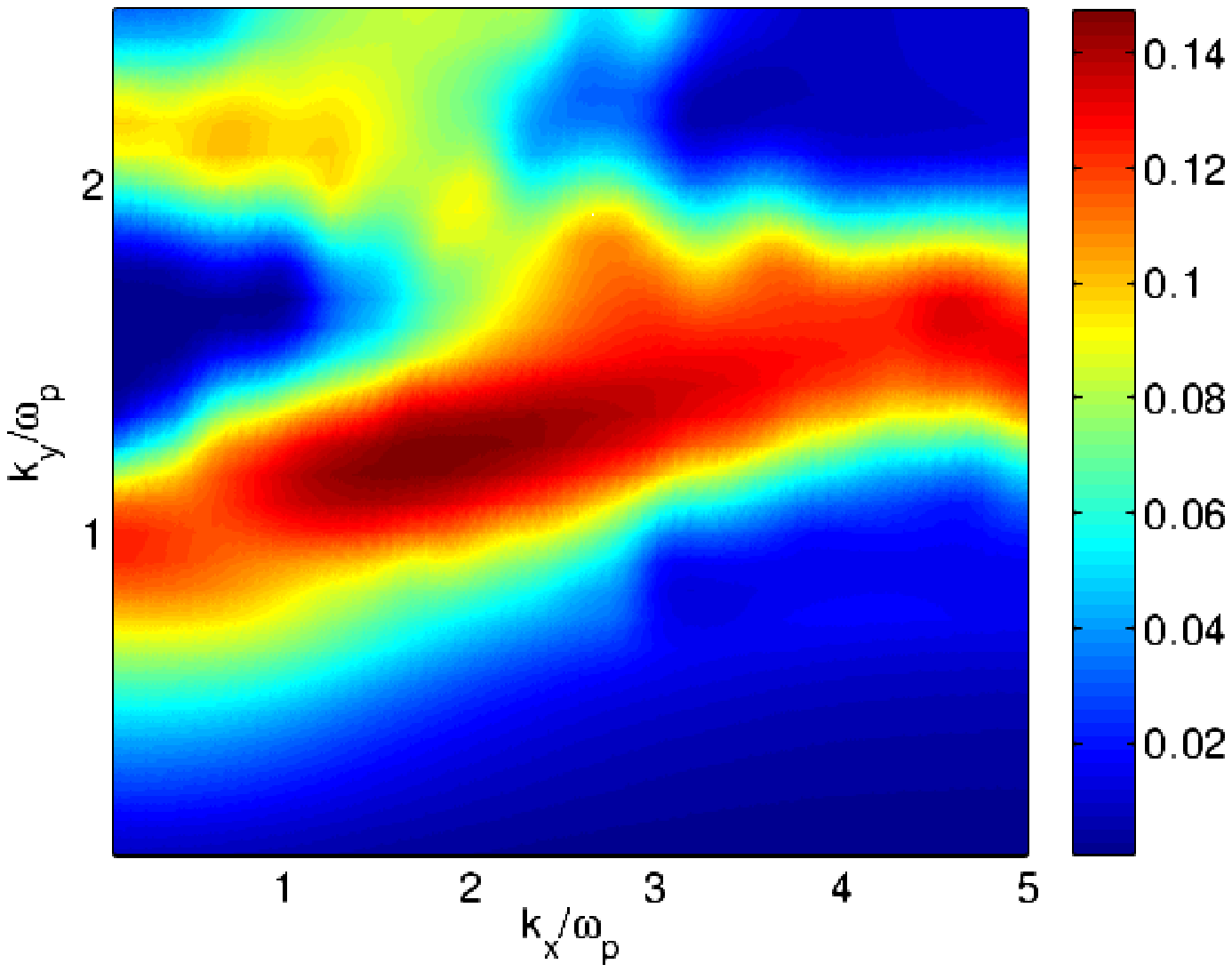}
              \justifying   \caption{ Growth rate map of the 2D oblique mode driven instability for a inhomogeneous beam-plasma
                 at transverse beam temperature $T_{b0\perp}$=10 keV at $k_{s}=3\pi$
                 and $\varepsilon=0.2$.}  
                 \label{fig:in4}
         \end{figure}

When the inhomogeneity scale length is sharper than the skin depth, i.e. for the case of 
 $ 2\pi/k_{s} \leq c/\omega_{0}$, there is an overall reduction of the growth rates in the system in comparison to the homogeneous 
 case. Furthermore, with   increasing  amplitude of density inhomogeneity 
also the growth rate reduces in this regime. This has been illustrated in the 2-D plot of the growth rate in Fig.~\ref{fig:in3} and ~\ref{fig:in4}.

 A detailed comparison of the  growth rate $\Gamma_{gr}$ of the maximally growing mode for various different values 
 of $k_s$ and $\varepsilon$ has been provided  
   in TABLE I.

\begin{center}
{\bf{TABLE I}} \\
The maximum growth rate of oblique mode driven instability evaluated analytically under the approximation of weak inhomogeneity amplitude as well as from PIC simulation. 
\vspace{0.2in}
\vspace{0.2in}
\begin{tabular}{c c c c c c c c c c c c c  ll}
\hline
\hline
       & $\varepsilon$  \hspace{0.3in}   &$k_{s}$ \hspace{0.3in}     &$\Gamma_{gr}$(max.)\hspace{0.3in} &$\Gamma_{pic}$(max.)\\
 \hline
      & 0.0  \hspace{0.3in}    &   0.0   \hspace{0.3in}      &0.1822    &0.1823\\
         & 0.1 \hspace{0.3in}     & $\pi$  \hspace{0.3in}      &0.1857  &0.1827\\
         & 0.1  \hspace{0.3in}    & $2\pi$  \hspace{0.3in}    &0.1811     &0.1818\\
      & 0.1  \hspace{0.3in}    & $3\pi$  \hspace{0.3in}    &0.1728     &0.1813\\
     & 0.2   \hspace{0.3in}   & $\pi$   \hspace{0.3in}      &0.1884  &0.1830\\
      & 0.2   \hspace{0.3in}   & $3\pi$   \hspace{0.3in}      &0.1475  &0.1659\\
     & 0.4  \hspace{0.3in}    &  $3\pi$  \hspace{0.3in}    &  0.1352 &0.16\\ 
\hline
\end{tabular}
  \end{center} 
  
  It is clear from the results that the growth rate drops when the amplitude as well as the wave number 
  of the plasma density inhomogeneity 
  is increased.  
\section{PIC Simulation}
In this section, we present the results from the $2\frac{1}{2}$D PIC simulations employed for the study of   beam plasma 
instabilities for an inhomogeneous density plasma. 
   The simulation box is a 
 Cartesian $X \times Y$
plane with periodic boundary condition for both the electromagnetic field and
charged particles.  The initial choice of configuration satisfies the equilibrium condition.  
The  electron beam is chosen to propagate in the $\hat{y}$ direction with a relativistic velocity $v_{0b}$ whose current is 
neutralized  by the cold return shielding current from the 
 background plasma electron  flowing in the  opposite  direction   with a velocity 
$v_{0p}$.  The respective densities are appropriately chosen for the current to be zero. 
Charge neutrality and a null current
density are both ensured  initially ($t=0$).  Thus, the system is field free  and in
equilibrium initially.

The electron beam has also been chosen to have a  finite temperature $T_{0b}$.  For the inhomogeneous plasma 
the pressure contribution in such cases have been  avoided by choosing $T_{0b}$ to be space dependent defined by  
eq.~{\ref{pressure_balance}}. 
The ions  are kept at rest during the simulation. The uniform plasma density $n_{0e}$ is taken as $10n_{c}$ where 
$n_{c}=1.1\times 10^{21} cm^{-3}$ is the critical density for 1$\mu$m wavelength of laser light. 
The area of the simulation box R is $30\times15$ $( c/ \omega_{0})^2$  
corresponding to $1500\times750$ cells where $
c/ \omega_{0}= d_{e}=5.0\times 10^{-2}\mu$m is the skin depth. 
The total number of particles is $45000000$ each for both 
electrons and ions. 
To resolve the underlying physics at the scale which is smaller than the skin depth, we have chosen
a grid size  of 0.02$d_{e}$. The time step is decided by the Courant condition. The time evolution of the box
averaged field energy density for every component of $\vec{E}$ and $\vec{B}$ are recorded at each time step. The 
energy density is normalized with respect to $(m_{e}c \omega_{0} /e)^2$ and 
time is normalized with respect to the electron plasma frequency $\omega_{0}$. 
 The results obtained from the simulation of the beam-plasma system for various amplitudes 
and scale lengths of inhomogeneity have been compared with the results of the homogeneous beam-plasma case. 
%
 The initial choices of the parameters for  simulation 
 are taken as $n_{0b}/n_{0p}=1/9$ or
$n_{0b}/n_{0e}=0.1$, $v_{0b}=0.9c$, $v_{0p}=-0.1c$ and $T_{0b\perp}=10 KeV$. 
These parameters favor the growth of the
oblique mode over the filamentation and two-stream modes.
 While considering the simulation of the  inhomogeneous case, the  inhomogeneity amplitudes $\varepsilon$ is varied from 
 $0.1$ to $0.2$. The inhomogeneity wavenumber is chosen 
 as  $k_{s}$=$\pi$, $2\pi$ and $3\pi$ for each of these amplitudes.
 
The perturbed field energies (magnetic as well as electric) are tracked in the simulation for the study of instabilities associated with the system. 
 The rate of exponential growth of the perturbed energy provides for twice  the growth rate of the 
  fastest growing mode associated with the instability. The growth rates $\Gamma_{pic}$(max.), thus evaluated for different cases of simulations 
  have been presented in one column of  TABLE I.
  
\begin{figure}[!htb]
        \centering
                \includegraphics[width=0.85\textwidth]{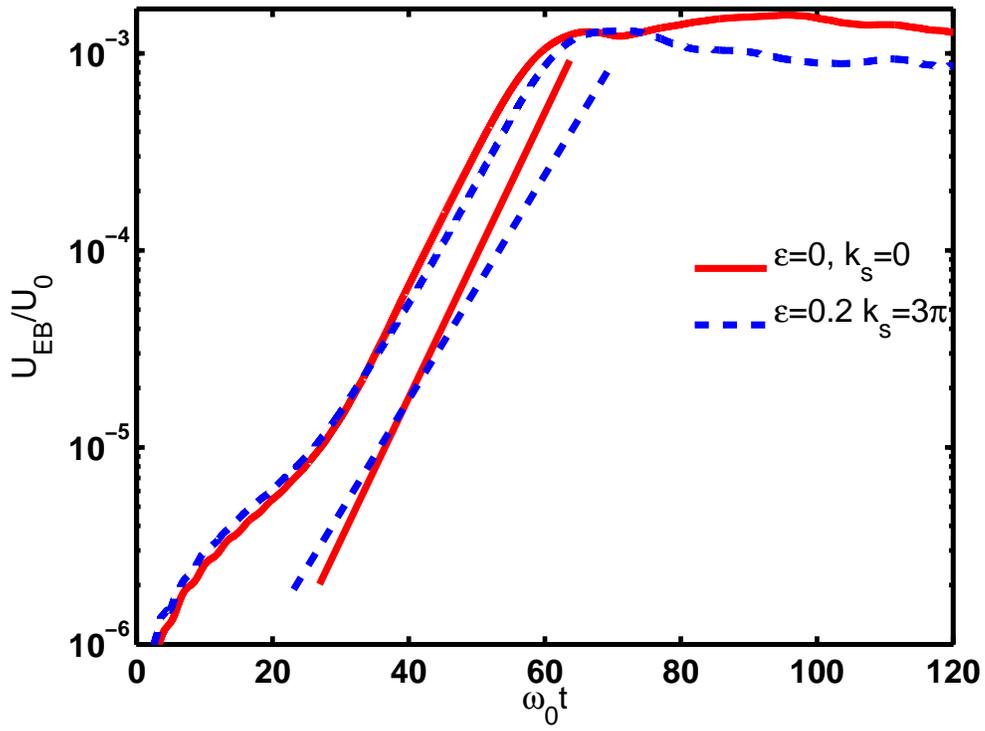}
\caption{ Growth rate from PIC simulation by energy slope.
                }
                 \label{fig:growth}
              \end{figure}

  The linear regime can be clearly 
  identified from a log plot of total energy shown in (Fig.~\ref{fig:growth}). The region of constant slope after an initial transient provides the growth 
  rate of the fastest growing mode.  The straight line alongside represents the slope obtained from the linear theory provided in section II. 
   It can be observed that the agreement between simulation and the theoretical linear results are remarkably good. The comparison also shows 
  that the reduction in the growth rate in the presence of inhomogeneity in plasma density for scales sharper than the skin depth. 
  
\begin{figure}[!htb]
        \centering
        \includegraphics[width=\textwidth]{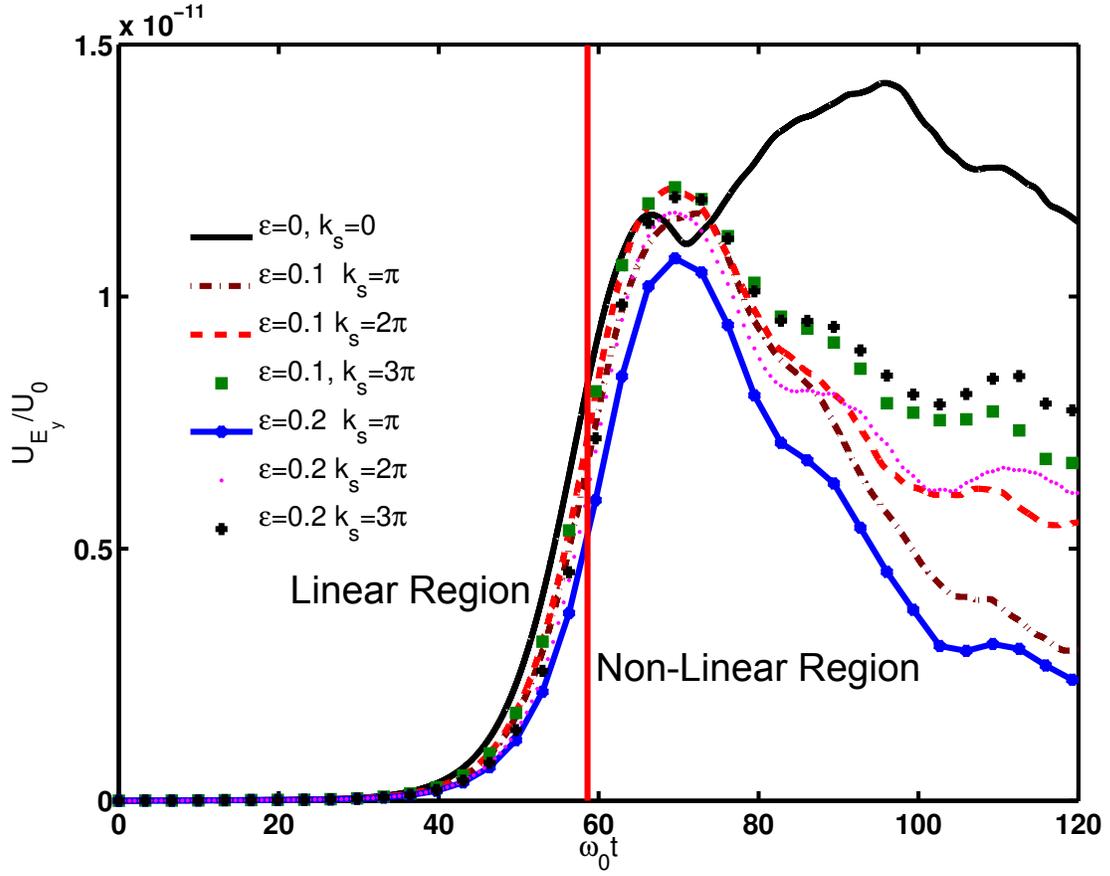}
 \caption { The evolution of y-component of normalized electric field energies $U_{E_{y}}$ with time for homogeneous and inhomogeneity cases. }
                  \label{fig:energy1}     
              \end{figure}

\begin{figure}[!htb]
        \centering
                 \includegraphics[width=\textwidth]{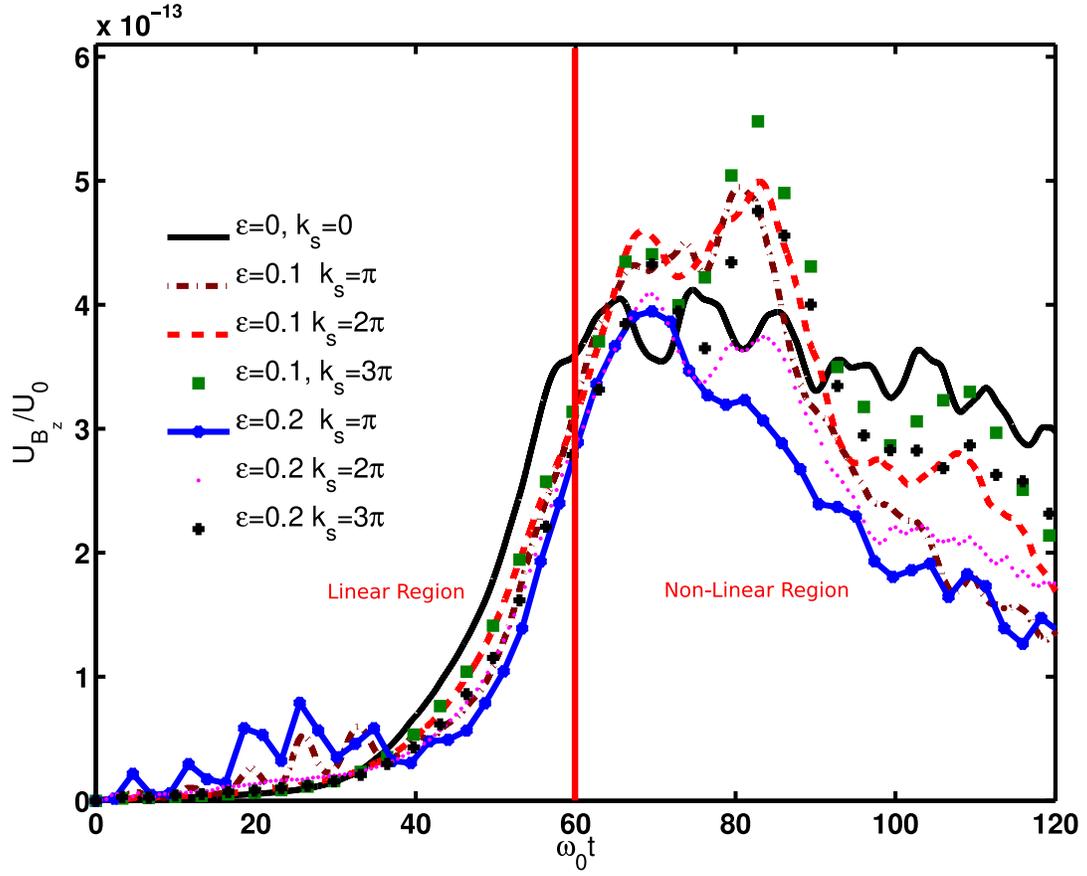}
\caption { The evolution of z-component of normalized magnetic field energies $U_{B_{z}}$ with time for homogeneous and inhomogeneity cases.  }
                        \label{fig:energy2}
                    \end{figure}

  The separate  evolution of electric field energy density  and the magnetic field energy density
 for various simulation cases 
  have been  shown in Fig.~\ref{fig:energy1} and Fig.~\ref{fig:energy2} respectively.  The energy associated with 
  electric field is observed to be typically always higher than the energy in magnetic field.
   It is interesting to compare these perturbed energies in the nonlinear regime. While for the homogeneous case 
  the  energy continues to remain high, there is a perceptible drop in both electric and magnetic field the energies 
  in the presence of inhomogeneity in the nonlinear regime.  This has important significance as it suggests that 
    the forward and reverse currents 
  which got separated during the linear phase have a tendency to merge again for the inhomogeneous 
  density  when nonlinearity sets in the system.

\begin{figure}[!htb]
        \centering
                \includegraphics[width=\textwidth]{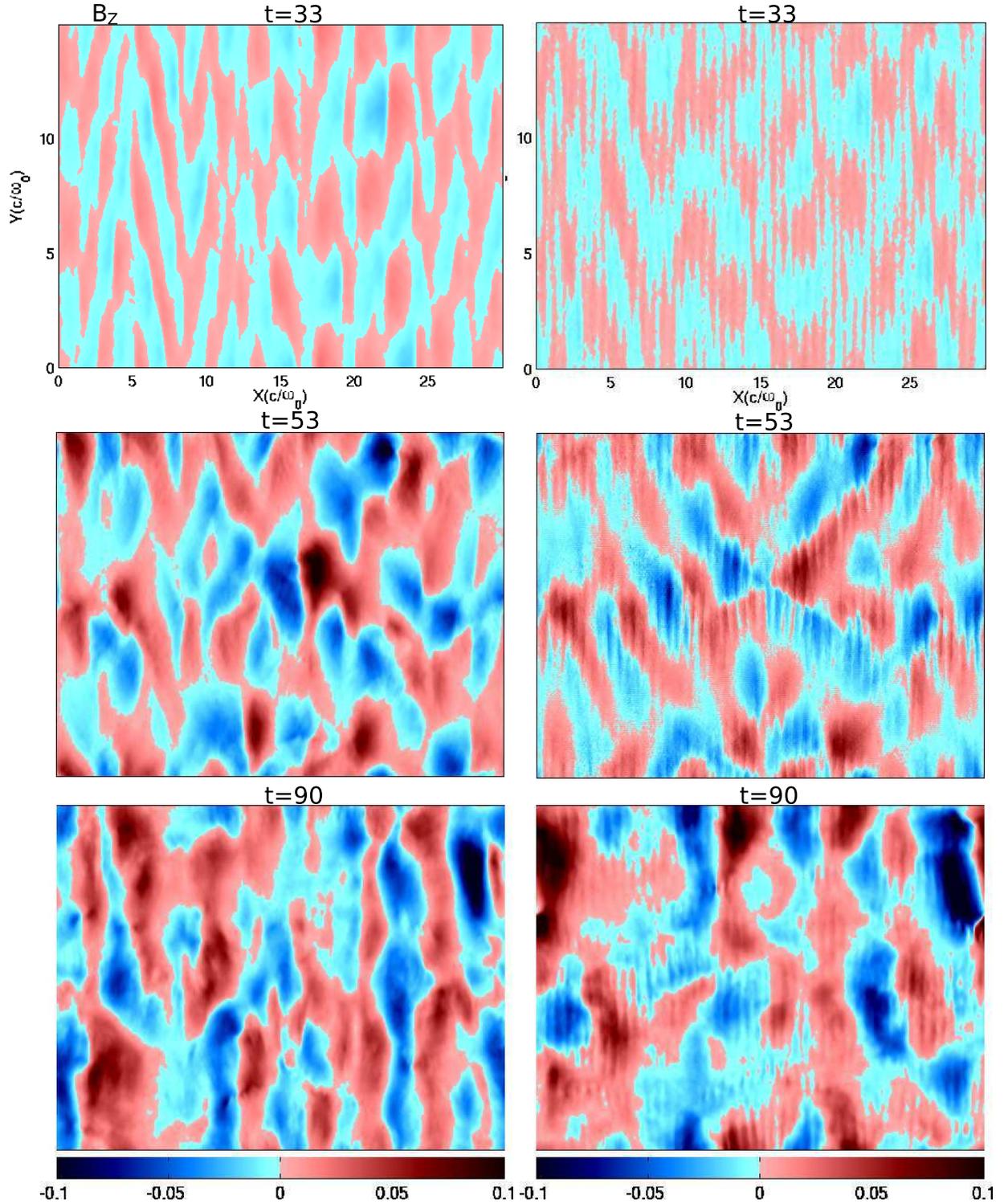} 
                 \caption { Time evolution and spatial configuration of z-component of magnetic field for homogeneous (first column) and inhomogeneous ($\varepsilon$=0.2 at $k_{s}$=$3\pi$, last column) case at time
         $\omega_{0}t=33$, $\omega_{0}t=53$ and $\omega_{0}t=90$.
         }
               \label{fig:mag}
              \end{figure}

\begin{figure}[!htb]
        \centering
                \includegraphics[width=\textwidth]{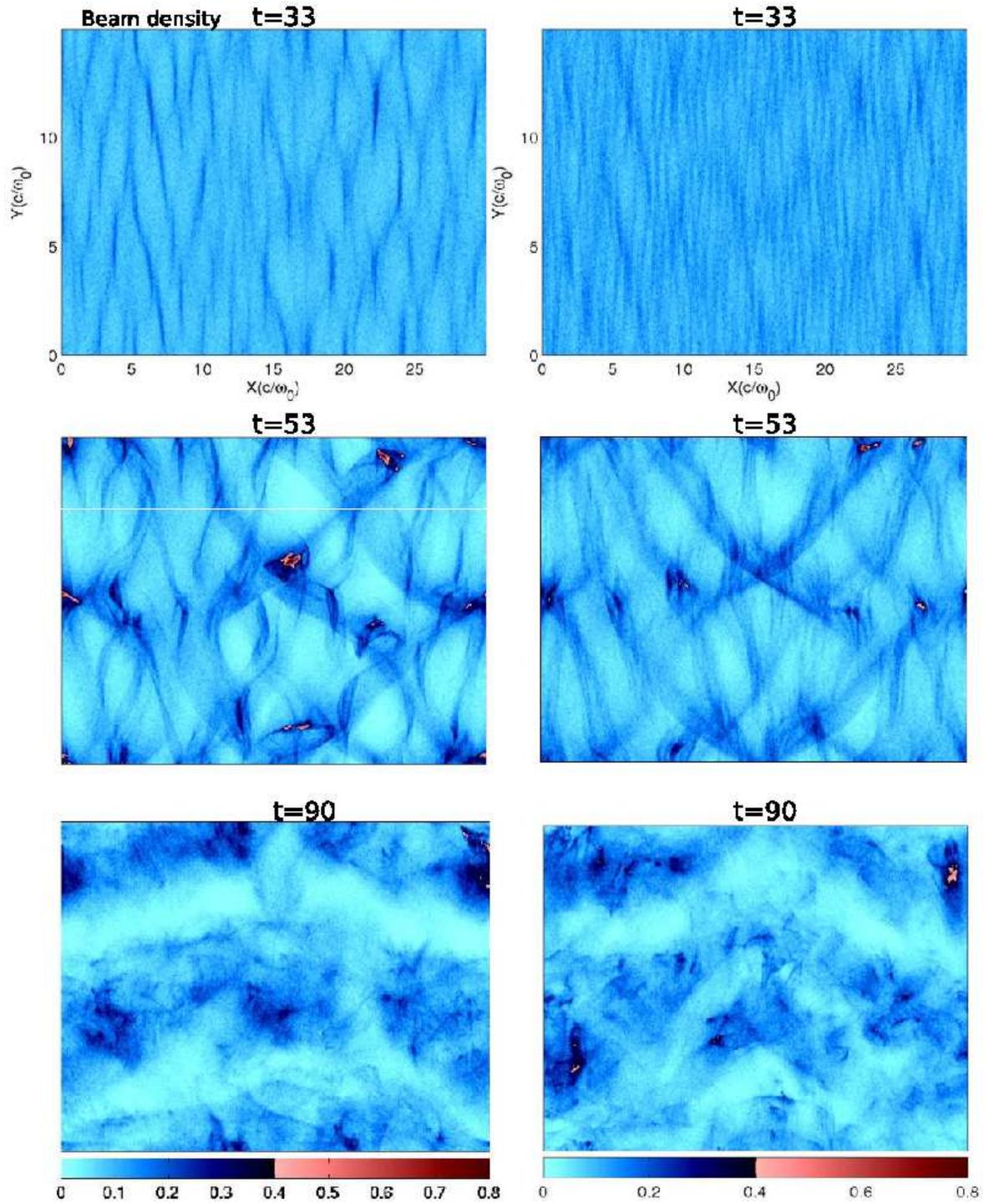}                 
                \caption { Time evolution and spatial configuration of beam density for homogeneous (first column)and inhomogeneous ($\varepsilon$=0.2 at $k_{s}$=$3\pi$, last column) case at time
         $\omega_{0}t=33$, $\omega_{0}t=53$ and $\omega_{0}t=90$.
         }
               \label{fig:beam}
              \end{figure}

\begin{figure}[!htb]
        \centering
                \includegraphics[width=\textwidth]{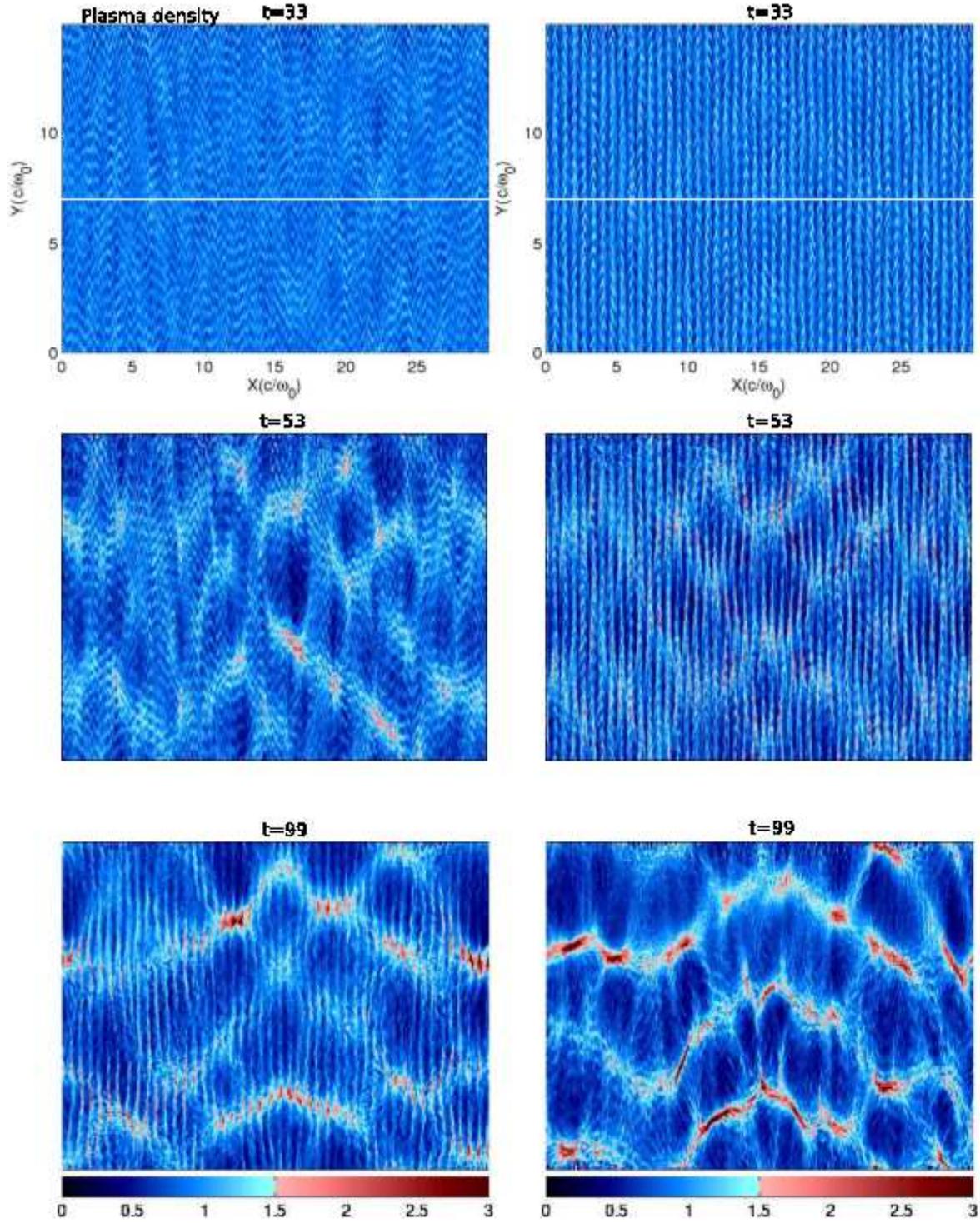} 
                 \caption { Time evolution and spatial configuration of background electron plasma density for homogeneous (first column)and inhomogeneous ($\varepsilon$=0.2 at $k_{s}$=$3\pi$, last column) case at time
        $\omega_{0}t=33$, $\omega_{0}t=53$ and $\omega_{0}t=90$.
         }
               \label{fig:background}
        \end{figure}

  The snapshot of the spatial profile in the $x-y$ plane of 
  certain  fields ($B_s$, $n_b$ and $n_p$)  at a time $\omega_{0}t=$33, 53 and 90 
  have been shown in Figs.~\ref{fig:mag}, \ref{fig:beam} and \ref{fig:background} respectively.  In these figures the first column corresponds to the homogeneous case, i.e. $\varepsilon = 0.0$ 
  and second column to the inhomogeneous case with $\varepsilon = 0.2 $ and $k_s = 3\pi$ (corresponding to density inhomogeneity 
  scale lengths to be sharper than the skin depth).  It is clear from these figures that the variations in these fields are in both $'x'$ as well as $'y'$ 
  directions, confirming that the oblique mode continues to dominate the instability scene. 
  However,  it should also be noted  that during the linear phase the magnetic field 
  acquires structures which are extended and aligned along the beam flow direction of $\hat{y}$. The structure size along the $x$ 
  direction is significantly  shorter and is found to compare with the typical values of the skin depth.  In the inhomogeneous case additional 
  structures  at the shorter inhomogeneity scale are observed to ride over those  appearing at the   skin depth scale.
  As an aside we wish to mention that fluid simulations based on Electron Magnetohydrodynamic model has also 
  shown that plasma density inhomogeneity leads to the guiding of electron current structures \cite{sharad}.
 These studies, therefore, suggest that specially tailored targets incorporating appropriate forms of plasma density 
 inhomogeneity can in fact be helpful for efficient transport of electron beam through plasmas. 
\section{Summary}
In this work instability of the beam plasma system has been analyzed for the case when the background plasma is 
chosen to be inhomogeneous. The linear analytical evaluation of growth rate shows that when the inhomogeneity scale length is shorter 
than the typical skin depth scale of the plasma the growth rate of the instability is suppressed. 
A quantitative comparison of the growth rate of the maximally growing mode for the homogeneous 
and inhomogeneous cases were made, which clearly show the reduction in the growth rate in the presence of density inhomogeneity 
with sharper scales. A detailed PIC simulation in 2-D have also been performed which confirms this. The simulation shows reduced growth rate 
and also the channelizing of current during the linear phase. 

Our work has direct relevance to the recent experimental work on 
the  observation of efficient transport of Mega Ampere of electron currents through
aligned carbon nanotube arrays. The ionization of the carbon nanotubes by the prepulse  of laser
pulse would produce an  inhomogeneous plasma density.  The suppression of the beam plasma instabilities 
would then aid the process of efficient transport of electron current as observed in the experiment.


\clearpage
\newpage
 \bibliographystyle{unsrt}

\FloatBarrier
%

\end{document}